\documentclass[aps,pra,twocolumn,amsmath,amssymb,showpacs,superscriptaddress,nobalancelastpage]{revtex4-1}

\usepackage{graphicx}% Include figure files
\usepackage{dcolumn}% Align table columns on decimal point
\usepackage[mathlines]{lineno}
\usepackage{bm}% bold math
\usepackage{graphicx}
\usepackage{xcolor}
\usepackage{url}
\usepackage[normalem]{ulem}
\usepackage{times}
\usepackage{txfonts}
\usepackage{bbm}
\usepackage{mathrsfs} %Fancy letters 
\usepackage{comment}
\usepackage{dsfont} %added by PJU
\usepackage[colorlinks=true,breaklinks=true,allcolors=blue]{hyperref}

\definecolor{jason}{rgb}{1,.2,.2}
\definecolor{jordan}{rgb}{.4,.2,1}
\definecolor{ben}{rgb}{.2,.6,.2}
\definecolor{andrew}{rgb}{0,.8,.5}
\definecolor{ebrahim}{rgb}{1,.5,0}

\newcommand{\grad}{\nabla}

\renewcommand{\d}{\partial}
\newcommand{\E}{\textbf{E}}
\newcommand{\B}{\textbf{B}}
\renewcommand{\S}{\textbf{S}}

\newcommand{\rhat}{\hat{r}}
\newcommand{\phat}{\hat{\phi}}
\newcommand{\zhat}{\hat{z}}

\newcommand{\me}{\mathrm{e}}
\newcommand{\ii}{i\mkern1mu}
\newcommand{\jj}{\mathbf{j}}
\newcommand{\sgn}{\text{sgn}}
\newcommand{\ket}[1]{\vert #1 \rangle}

\newcommand{\Sc}{Schr\"odinger~}

\begin{document}
%%%%%%%%%%%%%%%%%%%%%%%%%%% PREAMBLE %%%%%%%%%%%%%%%%%%%%%%%%%%%
\title{Non-radiating angularly accelerating electron waves}

\author{Jordan~Pierce}
\email[author contributions: ]{These authors contributed equally to the work}
\affiliation{Department of Physics, University of Oregon, 1585 E 13th Avenue, Eugene, Oregon 97403, USA}
\author{Jason~Webster}
\email[author contributions: ]{These authors contributed equally to the work}
\affiliation{School of Physics, University of the Witwatersrand, Private Bag 3, Wits 2050, South Africa} 
\author{Hugo Larocque}
\affiliation{Department of Physics, University of Ottawa, 25 Templeton St., Ottawa, ON K1N 6N5, Canada}
\author{Ebrahim Karimi}
\affiliation{Department of Physics, University of Ottawa, 25 Templeton St., Ottawa, ON K1N 6N5, Canada}
\author{Benjamin~McMorran}
\affiliation{Department of Physics, University of Oregon, 1585 E 13th Avenue, Eugene, Oregon 97403, USA}
\author{Andrew~Forbes}
\email[Corresponding author: ]{andrew.forbes@wits.ac.za}
\affiliation{School of Physics, University of the Witwatersrand, Private Bag 3, Wits 2050, South Africa}

%%%%%%%%%%%%%%%%%%%%%%%%%%% END %%%%%%%%%%%%%%%%%%%%%%%%%%%

%\date{\today}

\begin{abstract}
\noindent Accelerating electrons are known to radiate electromagnetic waves, a property that is central to the concept of many devices, from antennas to synchrotrons. While the electrodynamics of accelerating charged particles is well understood, the same is not true for charged matter waves:  would a locally accelerating charged matter wave, like its particle counterpart, radiate? Here we construct a novel class of matter waves, angular accelerating electron waves, by superpositions of twisted electrons carrying orbital angular momentum.  We study the electrodynamic behaviour of such accelerating matter waves and reveal the generation of a solenoidal magnetic field in each component, and an accelerating electron wave that does not radiate.  These novel properties will have practical impact in spin flipping of qubits for quantum information processing, have been suggested for control of time dilation and length contraction, and raise fundamental questions as to the nature of wave-particle duality in the context of radiating charged matter.
\end{abstract}
%%%%%%%%%%%%%%%%%%%%%%%%%%% END %%%%%%%%%%%%%%%%%%%%%%%%%%%
\maketitle

\section{Introduction}

\noindent Charged particles are known to emit light when accelerating through a process called bremsstrahlung.  A prominent example of this is the aurora, in which electrons and protons orbiting with the Earth's magnetic field release colorful light visible at extreme latitudes. This same effect causes synchrotron radiation and limits the maximum speed in circular particle accelerators. All of these cases result from a charged particle interacting with an external electromagnetic field and accelerating. In free space devoid of external fields, charged particles are commonly held to propagate in straight lines. Here we investigate freely propagating electron matter waves that twist and undergo angular acceleration in free space.
    
Recent work with diffractive electron optics in transmission electron microscopes (TEMs) has allowed for the reliable creation and study of structured electron beams \cite{Harris2015aa}. One such structure is a freely-propagating electron matter wave with a helical phase $e^{\ii\ell\phi}$, where $\ell$ is an integer and $\phi$ is the azimuthal coordinate, winding about the azimuth, called an electron vortex \cite{lloydrev17,bliokhrev17}. The phase vortex has an associated quantized orbital angular momentum (OAM). Electron beams with OAM-carrying phase vortices result from a long tradition within physics dating back to Dirac or even earlier \cite{dirac_quantised_1931,mcmorran_origins_2017}. The first experimental demonstration of electron vortices was in 2010 by Uchida and Tanomora \cite{uchida2010generation}. Uchida and Tanomora employed a graphite staircase to approximate a spiral phase plate, which is commonly used in optics \cite{sueda2004laguerre}. Current work utilizing off-axis holograms within a TEM allows for a high degree of control over the structure of a diffracted electron beam \cite{verbeeck2010production, mcmorran2011electron, harvey2014efficient}.  Electron vortex beams have applications in numerous scientific studies, including potential magnetic monopole detection \cite{fukuhara_electron_1983, tonomura_applications_1987, beche2014magnetic}, measuring magnetic properties \cite{lloyd2012interaction,rusz2013boundaries,grillo_observation_2017}, atomic scale resolution techniques \cite{verbeeck2011atomic}, and magnetic dichroism experiments \cite{lloyd2012quantized}.
    
Freely propagating electrons do not radiate. The state can be described as a coherent superposition of skewed semi-classical trajectories that follow straight paths of constant momentum \cite{mcmorran2011electron}. While the phase of the beam is helical, this is not directly observable. The cylindrical probability distribution of such beams is azimuthally uniform, marked only by a central dark nodal line along the optical axis.  In contrast, electrons that follow helical paths do radiate. 

\begin{figure}[htbp]
	\centering
	\includegraphics[width=\linewidth]{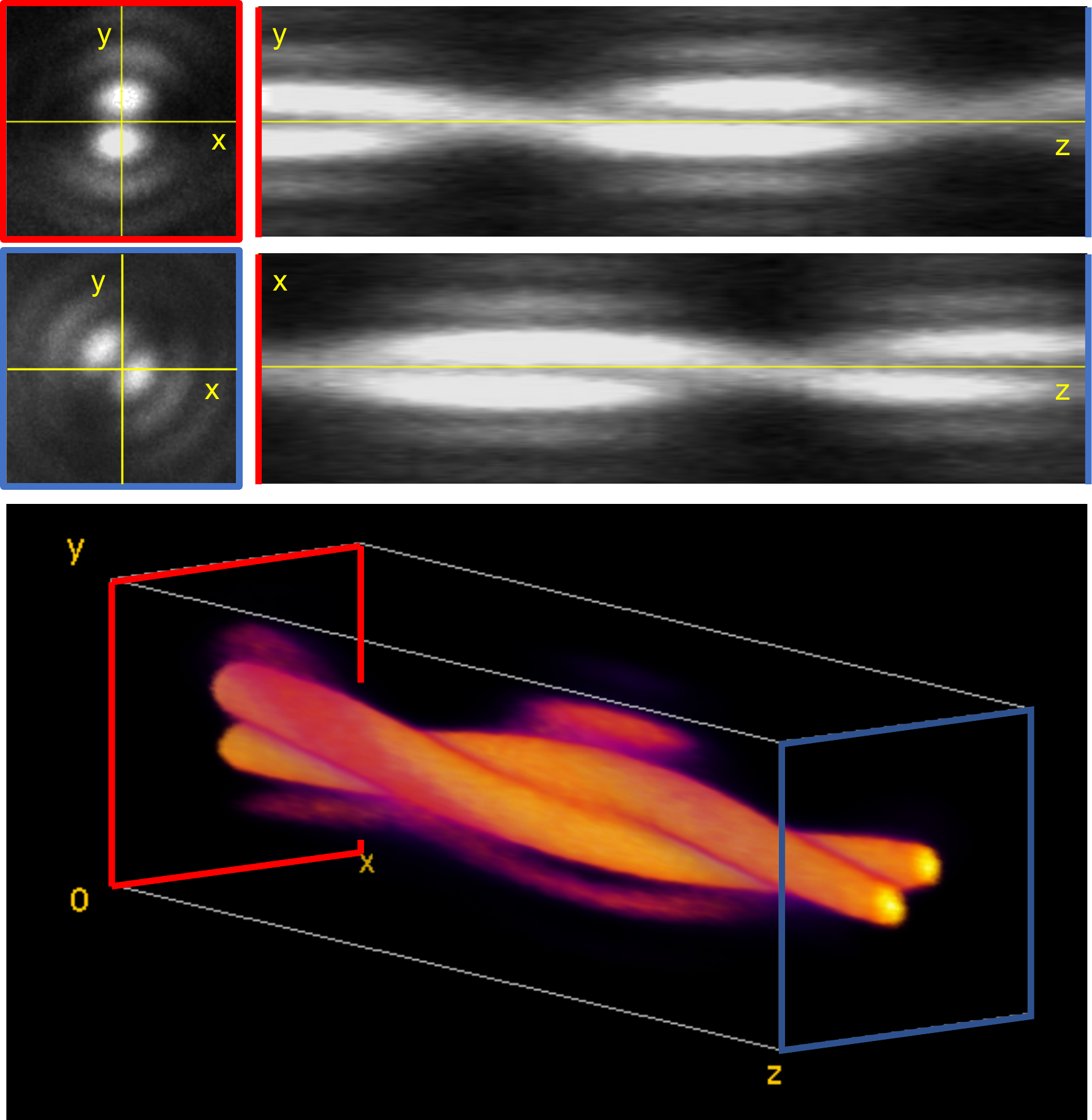}
	\caption{\textbf{A twisting electron beam.} Orthonormal projections of a 3D reconstruction of an experimental electron beam that twists in free space. The beam propagates in the +$z$-direction. The beam structure was determined by recording microscopic images of the beam in various $x$-$y$ planes along $z$ through a focal series, and then reconstructing the 3D structure using interpolation.}
	\label{fig:3Dtwist}
\end{figure}

Here we create electron matter waves that twist with a varying angular velocity as they propagate so that the wave function appears to follow a helical trajectory (see Fig.~\ref{fig:3Dtwist}). We achieve this using nanoscale holograms that create superpositions of OAM Bessel states \cite{grillo2014generation}.  The resulting electron beam is non-diffracting over a finite range, and the probability current locally accelerates azimuthally in a controllable manner.  We investigate the electrodynamic properties of these angularly accelerating electron beams and predict a solenoidal magnetic field in the component waves with zero radiation (zero Poynting vector) outward from the accelerating charged beam. A measurement of the electron energy spectra is consistent with a non-radiating state, despite local accelerations of the charged matter wave. That they do not radiate even while accelerating raises fundamental questions as to the nature of wave-particle duality in the context of charge.

\section{Theory}
\noindent First we outline the concept of our angularly accelerating matter waves and describe their propagation dynamics.  Next, we wish to understand such charged accelerating waves from an electrodynamics perspective, that is, determining the electric field $\E$ as well as the magnetic field $\B$ components, from which we can calculate the flow of energy in the form of electromagnetic waves from the charged matter wave. In doing so we wish to answer the question as to whether a charged matter wave will radiate due to internal acceleration as would happen due to external acceleration.

    The connection between optical beams and electron matter waves lies in the correspondence between the paraxial wave equation and the \Sc equation.  One solution to the Schr\"odinger wave equation in cylindrical coordinates $(r,\phi,z)$ for an electron matter wave function, ignoring spin, is given by the Bessel beam,  
    \begin{equation}
        \psi(r, \phi, z) = J_\ell(k_r r)e^{i\ell\phi} e^{ik_zz} \label{eq:Bessel}
    \end{equation}
    where $J_\ell$ is the Bessel function of order $\ell$; the wavenumber $k=\sqrt{2mE}/\hbar$ is related to the radial and longitudinal wavenumbers ($k_r$ and $k_z$ respectively) by the relation $k^2=k_r^2+k_z^2$, and the topological charge $\ell$ gives rise to an OAM of $\ell \hbar$ per electron. For a single Bessel mode, the phase varies linearly by $2\pi\ell$ about the azimuth; however, non-linear azimuthal phase profiles have been shown to allow the angular velocity of light waves to be varied \cite{schulze2015accelerated, webster2017radially}.

\begin{figure}[htbp]
	\centering
	\includegraphics[width=\linewidth]{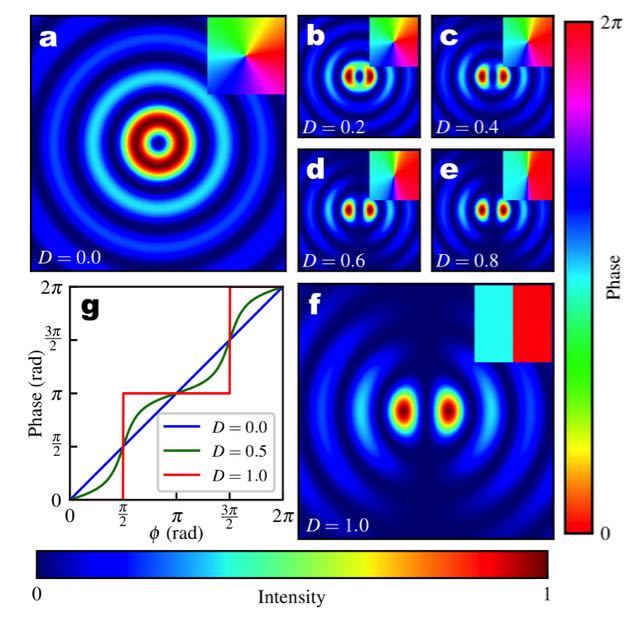}
	\caption{\textbf{Bessel superpositions.} Intensity and phase (insets) profiles of the matter wave for various values of the anisotropic parameter $D$. (a) Single Bessel beam with a linear azimuthal phase when $D = 0$, while (b)-(f) show profiles with increasing $D$. The azimuthal phase profile is shown in (g) where the non-linearity increases with $D$.}
	\label{fig:phase_unwrap}
\end{figure} 

\begin{figure}[htbp]
	\centering
	\includegraphics[width=\linewidth]{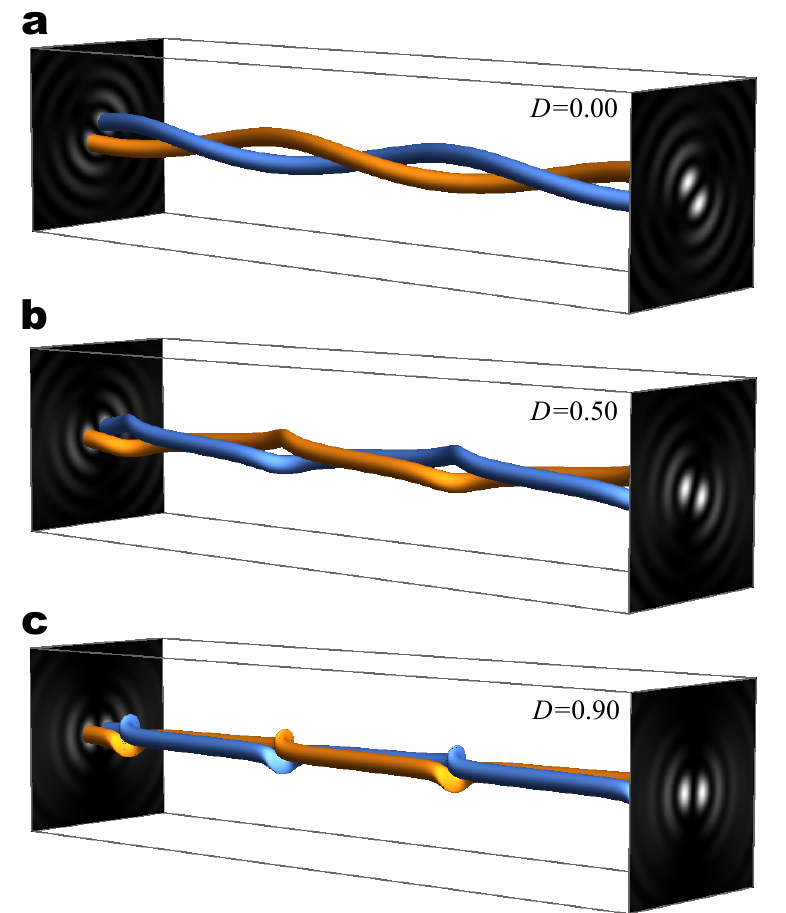}
	\caption{\textbf{Angular rotating waves.} Three dimensional representations of the angular rotation upon propagation of the electron waves as described by Eqs. \eqref{eq:sup_acc}.  (a) When $D = 0$ the wave rotates during propagation at a constant angular velocity, while for increasing $D$ the wave accelerates and decelerates, tightening the twisted motion, as shown in (b) and (c). The coloured tubes track the locations of the inner petal-like structure of the wave, i.e., the two ``intensity lobes'' shown in Figs.~\ref{fig:3Dtwist} and \ref{fig:phase_unwrap}.}
	\label{fig:twists}
\end{figure}

This prompts us to transform $\phi \to f(\phi)$ in \eqref{eq:Bessel} and then re-solve the free Schr\"odinger equation for $f(\phi)$. Doing so yields a new anisotropic expression for the wave function, given by
\begin{equation}
\psi(r,\phi,z) = J_\ell(k_r r)\left(1+\frac{2 D \cos(2\ell \phi)}{1+D^2}\right)^{\frac{1}{2}} \me^{\ii k_z z} \me^{\ii \ell \varphi_\ell(\phi)},
\label{eq:Bess_acc}
\end{equation}
where the non-linear phase profile is given by
\begin{equation}
\varphi_\ell(\phi)=-\phi+\frac{1}{\ell}\arctan\left(\frac{\sin(2\ell \phi)}{\cos(2\ell \phi)+D}\right),
\label{eq:varphi}
\end{equation}
and is plotted in Fig.~\ref{fig:phase_unwrap}(g). The new $D$ parameter which we referred to as the anisotropic parameter, takes any value between $0\leq D \leq 1$. It can also be shown (see Supplementary Information) that $D$ is a mixing parameter between two isotropic beams of opposite OAM states, where $D = 0$ is a single OAM state and $D = 1$ is an equal-amplitude superposition state. Increasing $D$ also affects the probability distribution of the matter wave, as shown in Fig.~\ref{fig:phase_unwrap}.

To generate an accelerating matter wave we must take a superposition of two anisotropic wave states with two differing longitudinal wave vectors $k_{z1}$ and $k_{z2}$, where the two waves also have opposite topological charges. We assume that the entire wave function is monoenergetic, and so the $k$ vector is the same within these two waves. The resulting superposition simplifies to
\begin{align}
\psi_{acc}(r,\phi,z)&=\left(1+\frac{2 D \cos(2\ell \phi)}{1+D^2}\right)^{\frac{1}{2}}\me^{\ii \bar{k}_z z} \nonumber \\
&\times \Big [ \mathcal{J}^+_\ell(r)\cos(\Delta k_z z+ \ell\varphi_\ell(\phi)) \nonumber \\
&+\ii\mathcal{J}^-_\ell(r)\sin(\Delta k_z z+ \ell\varphi_\ell(\phi))\Big ],
\label{eq:sup_acc}
\end{align}
where
\begin{align}
\mathcal{J}^\pm_\ell(r)&=J_\ell(k_{r1} r)\pm J_\ell(k_{r2} r), \\
\bar{k}_z&=\frac{k_{z1}+k_{z2}}{2}, \\
\Delta k_z&=\frac{k_{z1}-k_{z2}}{2},
\end{align}
which is an angularly accelerating and decelerating wave, consisting of $2\ell$ rotationally symmetric petal structures as depicted in Fig.~\ref{fig:twists}. To show that this field accelerates, we set the phase terms $\Delta k_z z+ \ell\varphi_\ell(\phi)$ to a constant which in turn allows us to track any point of interest within our field, for example, one of the petals as tracked in Fig.~\ref{fig:twists}.  From this we find that the rotation $(\Phi)$, angular velocity $(\partial_z\Phi)$, and angular acceleration $(\partial^2_z\Phi)$ as a function of the propagation distance are given by
\begin{align}
\Phi(z)&=-\frac{1}{\ell}\arctan\left(\frac{1+D}{1-D}\tan(\Delta k_z z)\right), \label{eq:acc} \\
\partial_z \Phi(z)&=-\frac{\Delta k_z}{\ell}\frac{1-D^2}{1+D^2-2D\cos(2\Delta k_z z)}, \label{eq:vel}\\
\partial_z^2 \Phi(z)&=-\frac{\Delta k_z^2}{\ell}\frac{4D(D^2-1)\sin(2\Delta k_z z)}{\left(1+D^2-2D\cos(2\Delta k_z z)\right)^2}. 
\end{align}
This acceleration is shown graphically in Fig.~\ref{fig:twists}(b) and Fig.~\ref{fig:twists}(c).

\begin{figure}[htbp]
	\centering
	\includegraphics[width=\linewidth]{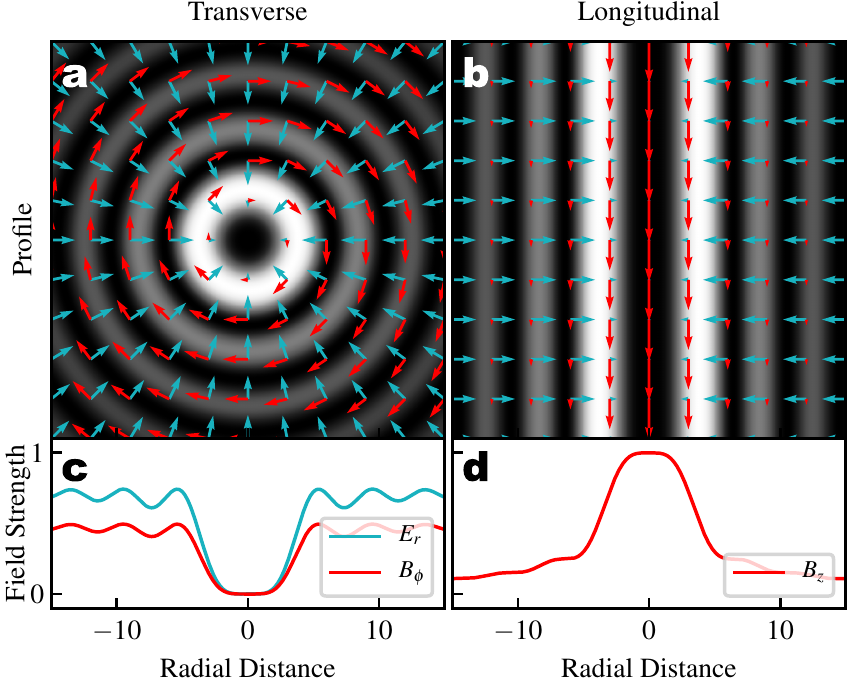}
	\caption{\textbf{Electric and magnetic fields.} Electric (blue) and magnetic (red) field components overlaid on the probability distribution (grayscale) for a non-accelerating isotropic electron Bessel beam in the (a) $xy$ and (b) $xz$ (b) planes. The corresponding electromagnetic field strengths are plotted in (c) and (d) in units where $k=1$ and $k_z=0.6$. The field strengths have been arbitrarily scaled to order one such that the fields can be meaningfully compared.}
	\label{fig:em}
\end{figure}

To calculate the $\E$ and $\B$ fields of the individual components of our wave, as given by \eqref{eq:Bessel}, we assume a probability density and current given by
\begin{gather}
\rho(r) = |J_\ell(k_r r)|^2, \label{eq:rho} \\
\jj (r) = \frac{\hbar}{m}|J_\ell(k_r r)|^2\left(\frac{\ell}{r}\phat+k_z\zhat\right), \label{eq:j}
\end{gather}
which are substituted into Maxwell's time-independent equations as the electric charge density and charge current, respectively. The analytic solution (see Appendix \ref{subsec:eBesselEM}) is shown graphically in Fig.~\ref{fig:em}.

Intriguingly, a solenoidal magnetic field $B_z$ arises in the centre of our field due solely to the OAM in our beam. Even more surprising, it can be shown that the $B_z$ field takes on the same value at the centre of the field regardless of the OAM. However, measuring such a field remains an open task, as the order of the magnetic field strength is around $10^{-8}$T, an extremely difficult quantity to observe within the cavity of a TEM.

Next we consider the radiating nature of such beams by calculating the radially outward component $\S\cdot\rhat=S_r$ of the Poynting vector $\S=(\E\times\B)/\mu_0$.  Radiation occurs if this outward component is non-zero infinitely far from the beam.  An electron wavefunction described by a single Bessel mode has an electric field with only a radial vector component. Thus, electrons in pure OAM states do not radiate; $S_r$ is zero in all parts of the beam. 
%Furthermore, any coherent superposition of OAM states, such as the accelerating wave solutions demonstrated here, must also be non-radiating because electric and magnetic fields from various component wavefunctions add vectorially \he{(Is this true? I would expect this if the beam consisted of two electrons in different OAM states but not a single electron in a superposition of states. Sure, the wavefunctions of a superposition are added coherently. But since the electric field and the magnetic field depend on the probability and the current densities, which depend on $|{\psi}|^2$ and $\psi^*\boldsymbol{\nabla}\psi$, then the problem of finding the fields of a superposition does not become as simple as vectorially adding those of the modes in the superposition)}. 

The vectorial component of a wavefunction's probability current density can be attributed to the gradient of its phase. That is, for a wavefunction $\psi$, the corresponding current density satisfies $\mathbf{j} \propto \boldsymbol{\nabla}\text{arg}\,\psi$, where $\text{arg}\,\psi$ refers to the argument or the phase of the wavefunction~\cite{berry2008exact}. For waves that have a well-defined longitudinal momentum $p_z$, this gradient can be used to provide an indicator of the flux lines attributed to the wavefunction upon propagation. More specifically, one can obtain these lines by letting the wave's transverse coordinates ($r,\phi$) in the latter become functions of the longitudinal coordinate ($z$), i.e. $\boldsymbol{\nabla}\text{arg}\,\psi (r,\phi,z) \rightarrow \boldsymbol{\nabla}\text{arg}\,\psi (r(z),\phi(z),z)$. Integrating over this gradient then provides the trajectories attributed to the wavefunction's probability current density. 

\begin{figure}[htbp]
	\centering
	\includegraphics[width=\linewidth]{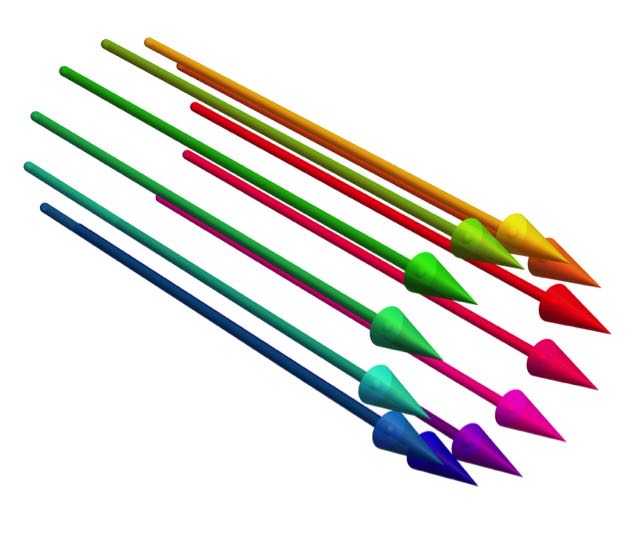}
	\caption{\textbf{Current lines of the wavefunction.} The fact that the probability flux lines are straight readily implies that the wavefunction's classical, or ray optics trajectories, are straight as well. The rays are coloured based on their initial azimuthal position and calculated following \eqref{eq:phaseAcc}.}
	\label{fig:trajectories}
\end{figure}

The phase of the angularly accelerating wavefunction presented in Eq.~(\ref{eq:sup_acc}) is given by:
\begin{equation}
	\label{eq:phaseAcc}
	\text{arg}\, \psi_{acc} = \arctan\left(\frac{J_\ell(k_{r1}r)-J_\ell(k_{r2}r)}{J_\ell(k_{r1}r)+J_\ell(k_{r2}r)} \tan(\Delta k_z z+\ell \varphi_\ell(\phi))\right).
\end{equation}
Given that we consider accelerating wavefunctions where $k_{r1} \approx k_{r2}$, then the above equation readily only yields values of $0$ and $\pi$. For this reason, we can expect the gradient of Eq.~(\ref{eq:phaseAcc}) to be zero and thus the current lines to correspond to straight trajectories as shown in Fig.~\ref{fig:trajectories}.  We thus hypothesize that neither the accelerating nor non-accelerating beams radiate.

\section{Results}

\begin{figure}
	\centering
	\includegraphics[width=\linewidth]{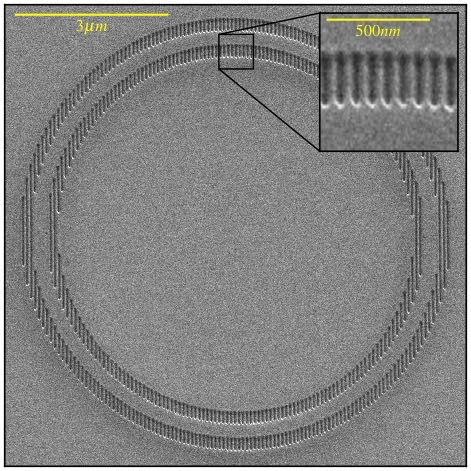}
	\caption{\textbf{Hologram.} Our hologram is placed in the far field and consists of a double ring slit of differing radii, producing an accelerating wave in the Fourier plane.}
	\label{fig:grating}
\end{figure}

\begin{figure*}[t]
	\centering
	\includegraphics[width=15cm]{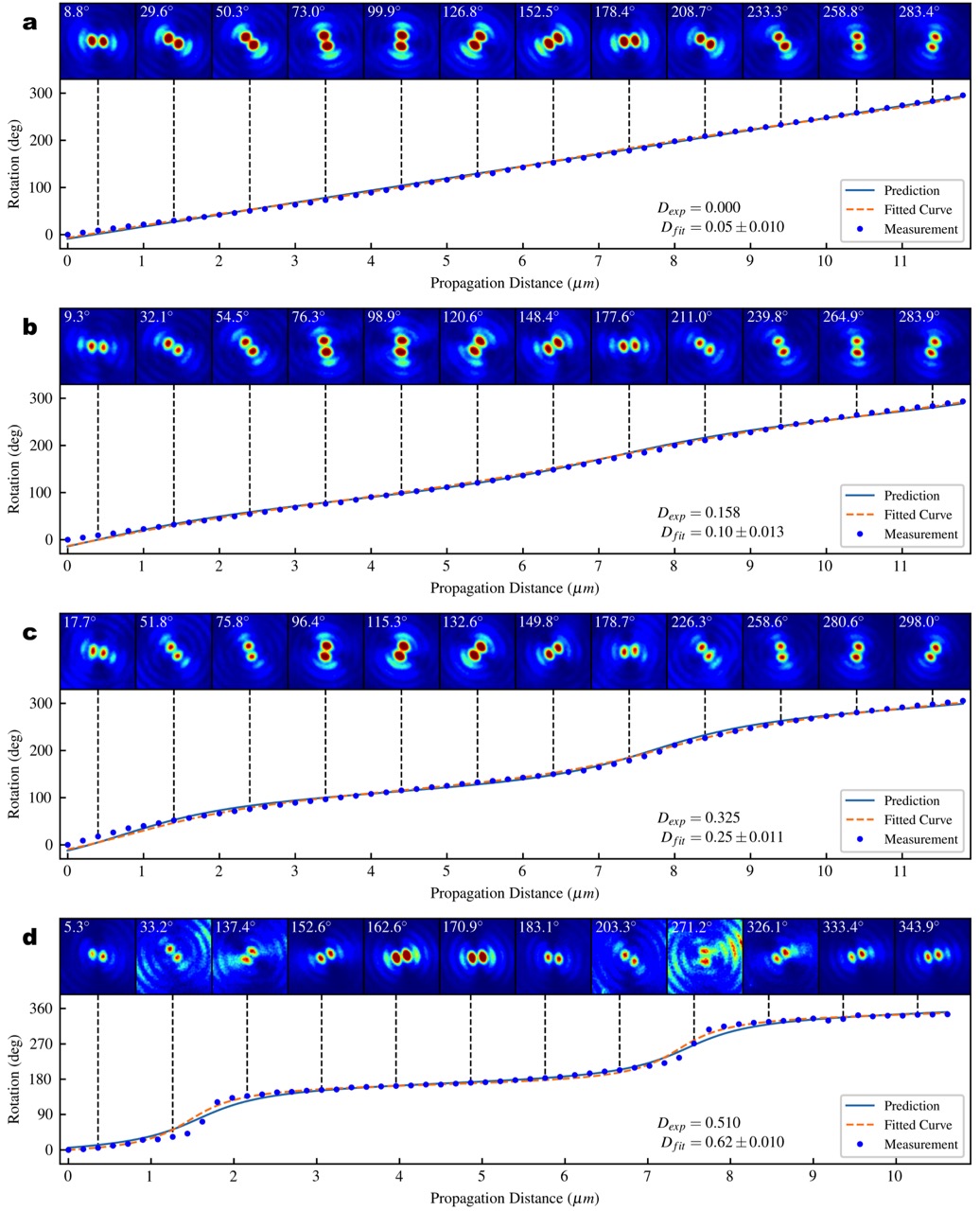}
	\caption{\textbf{Rotating matter waves.} Experimental results for the rotation of the matter waves as a function of propagation distance. $D_{exp}$ is the encoded anisotropic parameter on our hologram while $D_{fit}$ is the calculated parameter from the data.  The theoretical (blue curve), fitted (orange curve) and measured rotation (points) are plotted together for comparison for the cases of (a) $D_{exp} = 0$, (b) $D_{exp} = 0.158$, (c) $D_{exp} = 0.325$, and (d) $D_{exp} = 0.510$. The insets show the measured detected electron distributions with the white text the measured rotation of wave relative to the first frame in the sequence.}
	\label{fig:rots}
\end{figure*}
	
\textbf{Experiment.} We create our angular accelerating electron wave with a nanofabricated hologram placed in the front focal plane of a magnetic lens, such that the Fourier-transform of the hologram occurs in the back focal plane of the lens. The imaging system of the electron microscope demagnified and projected this image onto the detector. The focal conditions of this imaging system were adjusted such that the beam could be imaged at various locations along the optical axis. The Fourier transform $\mathcal{F}[\Psi]$ of the desired wave was encoded onto the hologram using an off-axis design. The Fourier transform of a single Bessel mode is $\mathcal{F}[\Psi(r)]=\me^{\ii \ell \phi}\delta(r-r')$, forming a ring of infinitesimal thickness we call a `delta' ring. The off-axis hologram design encodes the phase using a sinusoidal carrier. The carrier contains a fork dislocation \cite{mcmorran2011electron} in order to encode the OAM in our beam (see Fig.\ref{fig:lens}). The generated and manufactured holograms can be seen in Fig. \ref{fig:grating}.  The delta ring must have a finite thickness in order to host the sinusoidal carrier and diffract a reasonable electron current into the desired beam. However, the non-zero ring thickness approximating the radial delta function results in a diffracted beam of finite width, which affects the quality of the generated Bessel beam and the focal range over which it is approximately Bessel-like.  Thus we chose a width such that the diffracted beam was very close to a true Bessel beam within the experimental focal range, while still having sufficient intensity.  To create the twisting electron beam composed of superpositions of Bessel beams, the hologram was designed with two distinct delta rings with a slightly different $r'$ parameter.  The thickness of each ring is 200 nm, while the diameter of the rings is 8.0 $\mu$m and 7.0 $\mu$m.  To separate the diffracted beams, the period of the carrier wave is 75 nm.   The relatively small diameter of the holograms was chosen so that the resulting diffracted beam would be large enough in our detector to observe the detail of the beam.

\begin{figure}
	\centering
	\includegraphics[width=\linewidth]{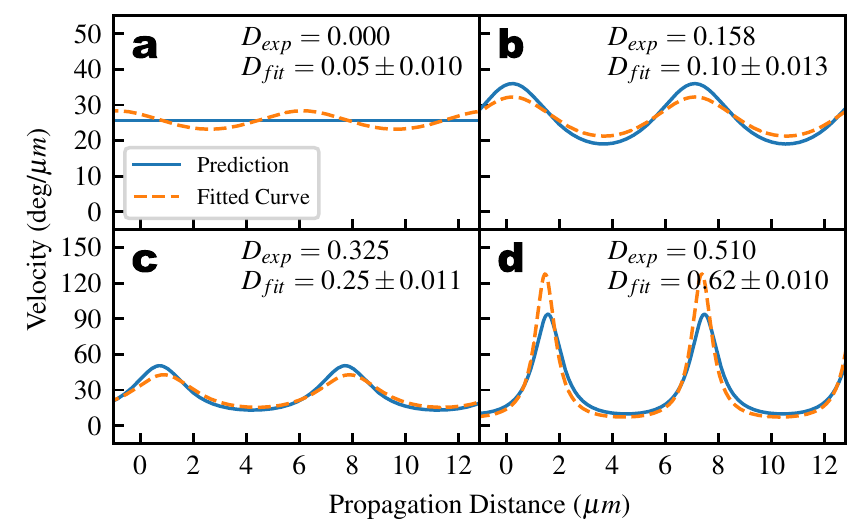}
	\caption{\textbf{Measured angular velocity.} The measured velocity of the wave as it propagates for $D_{exp} = 0$, (b) $D_{exp} = 0.158$, (c) $D_{exp} = 0.325$ and (d) $D_{exp} = 0.510$. The solid blue curve is the theoretical result based on the programmed hologram ($D_{exp}$) shown together with the measured data (points) and the fitted curve using $D_{fit}$ (orange dashed curved).}
	\label{fig:vel}
\end{figure}

\begin{figure}
	\centering
	\includegraphics[width=\linewidth]{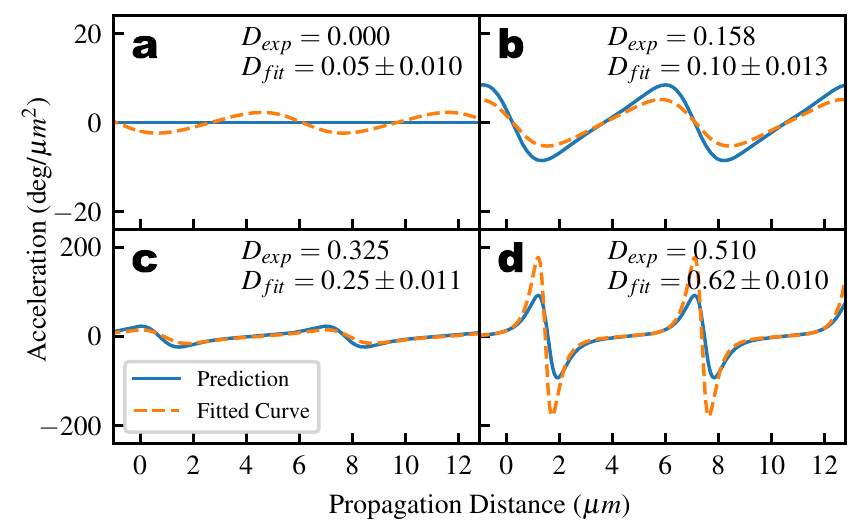}
	\caption{\textbf{Measured angular acceleration.} The measured acceleration of the wave as it propagates for $D_{exp} = 0$, (b) $D_{exp} = 0.158$, (c) $D_{exp} = 0.325$ and (d) $D_{exp} = 0.510$. The solid blue curve is the theoretical result based on the programmed hologram ($D_{exp}$) shown together with the measured data (points) and the fitted curve using $D_{fit}$ (orange dashed curved).}
	\label{fig:accs}
\end{figure}

\textbf{Rotation and acceleration measurements.}  Images of the beam's intensity profile were taken for $\ell=1$ and $D = 0$, $0.158$, $0.325$, and $0.510$ at equally spaced locations along the optical axis in a focal series (see Methods), with the results shown in Fig.~\ref{fig:rots}. The anisotropic parameter that we encoded into the diffraction grating is given by $D_{exp}$ in the figures and serves as the ``expected" anisotropic parameter, while $D_{fit}$ is retrieved by fitting \eqref{eq:acc} to the data and retrieving the $D$ parameter from this fit. 

The expected and fitted acceleration points align very closely. We note that small changes in $D$ make almost imperceptible differences in the rotation unless $D$ is close to one, as seen in the difference between the fitted and predicted curves in Fig.~\ref{fig:rots}. With these considerations in mind, we claim that the measured values given by $D_{fit}$ align well with the encoded $D_{exp}$.  The measured angular velocity and acceleration is given in Figs.~\ref{fig:vel} and \ref{fig:accs}. Our measured velocity and acceleration closely match with the predicted data, which is expected given that the fitted anisotropic $D$ parameters aligns quite closely to those that we predicted (programme onto the hologram).

\textbf{Radiation measurements.} Electron energy loss spectra (EELS) is a standard method for measuring the energy spectrum of the inelastic scattering events of the electron beam interacting with a sample.  The same method can also be used to measure any radiative events that the beam may undergo during acceleration.  The minimum energy loss measureable with a given EELS system depends on the energy spread of the source, and EELS data are unreliable at detecting energy loss within the width of the zero loss peak.  Our system should be able to measure 2 eV or above reliably.  
    
\begin{figure}
	\centering
	\includegraphics[width=\linewidth]{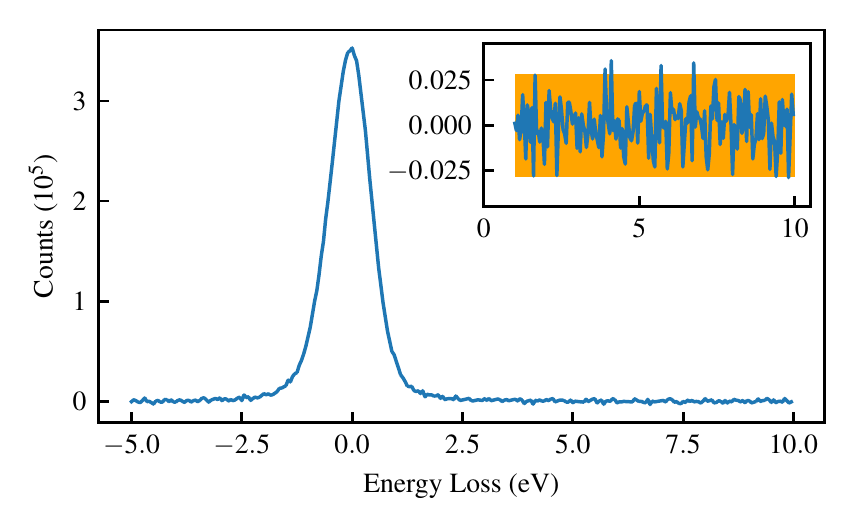}
    \caption{
        \textbf{Electron Energy Loss Spectra.} The EELS data for a $D=0.51$ beam.  Any radiative energy loss due to the acceleration of the beam should show as a peak at the corresponding energy loss value, whereas no such peak is observed.  The inset shows the signal of the accelerating beam subtracted from a reference beam that passes off-axis through the hologram (no acceleration encoded) where the orange band represents the 95$\%$ noise bounds.  The signal is entirely within the noise limits of the detector. This data is consistent with zero radiative energy loss.}  \label{fig:eels}
\end{figure}

Measured EELS data is shown in Fig. \ref{fig:eels} for $D=0.51$ as an example.  The microscope was in the same optical conditions during the EELS acquisition as during the data collected for the acceleration measurement images, and none of the beams showed any EELS signal apart from the zero loss peak (ZLP), being consistent with no radiation during the acceleration phase.  When the ZLP is subtracted the remaining signal is entirely within the noise limits of the detector, as seen in the inset.  %This is in stark contrast to what might be expected if the electrons themselves were following helical trajectories.  In this case the radiation would be significant and in a range measurable by EELS (see Supplementary Information).

\section{Discussion and Conclusion}
We have demonstrated a twisted electron beam with a rotating intensity distribution within the cavity of a TEM. The beam was analysed theoretically and predictions were made on the possible electromagnetic field that can arise from such a beam, which showed that both accelerating and non-accelerating beam profiles do not radiate. We also noted that a solenoidal magnetic field can arise from the centre of the non-accelerating beam due to its OAM, and such a field has not been measured previously. We then tested our theoretical predictions by first measuring the rotation of the field as it propagated, and the results showed a good agreement with the theoretical predictions. The angular velocity and acceleration of the field could be inferred and showed good agreement with the theoretical predictions. We can thus conclude with certainty that we have developed a method for controlling the angular acceleration of an electron's wavefunction in a predictable manner. We have also provided a number of open tasks to explore such as the measurement of a lack of radiation produced by our beams and the creation of a solenoidal magnetic field at the centre of these electron vortex beams.  We point out that while we have demonstrated this phenomenon with Bessel wave functions, it may also be executed with other suitable functions, e.g., Laguerre-Gaussian wave packets (see Supplementary Information).

Our work introduces a new class of matter waves with intriguing properties:  such structured accelerating waves have been suggested to have relevance to fundamental physical processes such as time dilation and length contraction \cite{kaminer2015aa}, while the solenoidal field would be a useful tool for spin-flipping operations in the context of quantum information processing.  The counter-intuitive notion that the accelerating wave does not radiate while a particle would is somewhat resolved by the realisation that the centre of mass $\langle x \rangle$ of the electron beam remains stationary. However, no electrons are ever expected to be found at this location as the beam has zero intensity along its centre. It's the probability distribution for an electron that informs us as to where we might find it, and this has been shown to be an accelerating field. Our intuition begins to break down again, as we see that we have somehow created a beam where the positions of the electrons accelerate whilst their centre of mass remains fixed. Our results thus reveal that the centre of mass of an electron has to accelerate in order to produce radiation.

%%%%%%%%%%%%%%%%%%%%%%%%%%% ACKNOWLEDGEMENTS %%%%%%%%%%%%%%%%%%%%%%%%%%%
\section*{Materials and correspondence}
Correspondence and requests for materials should be addressed to AF.
%%%%%%%%%%%%%%%%%%%%%%%%%%% END %%%%%%%%%%%%%%%%%%%%%%%%%%%

%%%%%%%%%%%%%%%%%%%%%%%%%%% ACKNOWLEDGEMENTS %%%%%%%%%%%%%%%%%%%%%%%%%%%

\section*{Acknowledgments}
AF acknowledges financial support from the South African National Research Foundation; BJM and JSP acknowledge support by the National Science Foundation under Grant No. 1607733. HL and EK acknowledge support from the Canada Research Chairs. The authors would like to thank Melanie McLaren, Elias Sideras-Haddad, and Vincenzo Grillo for useful discussions.
%%%%%%%%%%%%%%%%%%%%%%%%%%% END %%%%%%%%%%%%%%%%%%%%%%%%%%%

%%%%%%%%%%%%%%%%%%%%%%%%%% AUTHORS' CONTRIBUTIONS %%%%%%%%%%%%%%%%%%%%%%%%%%

\section*{Authors' contributions}
JP and BM performed the experiments with JW, AF, HL, and EK providing the theoretical analysis.  All authors contributed to the data analysis and the writing of the manuscript.

%%%%%%%%%%%%%%%%%%%%%%%%%%% END %%%%%%%%%%%%%%%%%%%%%%%%%%%

%%%%%%%%%%%%%%%%%%%%%%%%%%% FINANCIAL INTERESTS %%%%%%%%%%%%%%%%%%%%%%%%%%%

\section*{Competing financial interests}
The authors declare no financial competing interests.

\section{Methods}

\textbf{Hologram preparation.} The holograms were milled into a 75 nm thick Silicon Nitride membrane. Using e-beam evaporation, the membrane was coated on both sides with 5 nm Titanium followed by 10 nm Platinum.  This metal coating mitigates charge buildup within the membrane during milling and within the TEM.  The membrane was 250 $\mu m^2$ suspended on a Silicon base.  These membranes are industry standard TEM sample holders, are readily available in many thicknesses, and are ideal for electron holograms due to their rigidity, electron transparency properties, and because they are amorphous.  The holograms were milled into the membranes with a Helios 650 FEI Focused Ion Beam (FIB).  A set of 12 holograms were made, with three different $\ell$ values, and four anisotropic $D$ parameters.  For each value of $\ell$, the anisotropic parameters were $D=0.0$, $0.158$, $0.325$ and $0.510$.  The holograms were milled to a depth of between 30 and 35 nm, which corresponds to an approximate phase depth for 300 keV electrons in Si$_3$N$_4$ \cite{harvey2014efficient} This depth maximizes the diffracted efficiency of the hologram.

\textbf{TEM.} The electron diffraction experiments were performed in an FEI Titan TEM equipped with a spatially coherent field emission gun. The set of diffraction holograms was installed in the sample holder, and illuminated with a wide beam to maximize spatial coherence. The beams were viewed in low angle diffraction (LAD) mode, and to maximize the magnification, the effective camera length was set to 1350 m. In this mode, the objective aperture was used to select which hologram was being propagated through the system. The diffraction patterns were recorded in 15 second exposures using a Gatan Enfina imaging detector. We then collected focal series over a large enough defocus range to observe any angular acceleration.  The defocus range included a defocus of zero. The images were then aligned to account for lateral beam drift during the focal series.

\textbf{EELS data.} Using the same microscope conditions as during the image acquisition measurements, with the beams at focus the beams were propagated through an electron dispersion system to spread the beam based on electron energy.  Using a binning of 0.05 eV per pixel and an image width of 2048 pixels, the EELS data were recorded for each of the created holograms, and analyzed for any EELS signal.

\textbf{Fitting routines and errors.}  The rotation in the fields was measured by using a least squares fitting, where we fit each image to some preselected image of higher quality. All reported measurements have an error of less than $0.5$ degrees, which is the error within the least squares fitting alone. Other errors due to the experimental setup, such as any spherical aberration or errors within the manufacturing of the grating (errors that must be induced in order to create a ring slit, such as a non delta-like ring) are not accounted for in these measurements.  Though it is hard to perceive in Fig.~\ref{fig:rots}, our data still possess a small amount of noise, where each measurement point could deviate by approximately 2 degrees from it's true measurement. Due to this noise, we cannot simply take discrete derivatives with respect to the frames without amplifying this noise dramatically.  Instead, we make use of fitting \eqref{eq:acc} to our rotation data, before taking the second derivative of this fitted curve.  The curves as fitted in Fig.~\ref{fig:rots} are shown to match the data very well, and so our measured acceleration should not bear any significant biases in favour of our predictions.

%\bibliography{Bib}


\newpage

\section{Supplementary Information}

\subsection{The mixing parameter $D$}

We can label an isotropic state with an OAM of $\ell$, given in \eqref{eq:Bessel}, as $\ket{\ell}$. By performing a modal decomposition on the anisotropic state from \eqref{eq:Bess_acc}, labelled here as $\ket{\psi}$, we reveal that
\begin{equation}
\ket{\psi} = \frac{1}{\sqrt{1+D^2}}\ket{\ell}+\frac{D}{\sqrt{1+D^2}}\ket{-\ell}.
\end{equation}
From this we see that when $D=0$, we have $\ket{\psi}=\ket{\ell}$ - a single mode state. When $D=1$, we have $\ket{\psi}=(\ket{\ell}+\ket{-\ell})/\sqrt{2}$, which is a fully balanced superposition state between $\ket{\ell}$ and $\ket{-\ell}$. Values of $D$ that exceed one can be used, but this has an equivalent formulation when one reverses the topological charge $\ell\to -\ell$, thus $D>1$ doesn't contribute to the discussion.

\subsection{Generating Bessel wave functions}
\begin{figure}[h]
	\includegraphics[width=\linewidth]{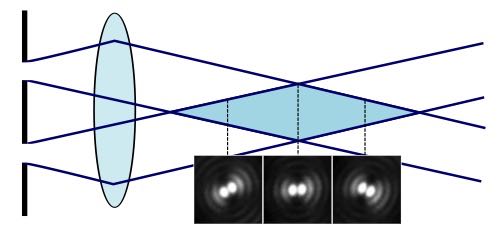}
	\caption{(a) Depiction of a ring slit and lens used to create a Bessel wave function that is valid over the shaded region.}
	\label{fig:lens}
\end{figure}

Our Bessel wave functions are created in the far field by exploiting annular rings as the Fourier transform to the desired function.  After Fourier transforming the annular ring, the Bessel wave function is created over a finite region, as depicted in Fig.~\ref{fig:lens}.

\subsection{The Helmholtz equation in quantum mechanics and its solutions}
The Schr\"{o}dinger Equation is given by
\begin{equation}
\left(-\frac{\hbar^2}{2m}\grad^2+V(\mathbf{r},t)\right)\Psi(\mathbf{r},t)=i\hbar \frac{\d}{\d t}\Psi(\mathbf{r},t),
\end{equation}
where $V(\mathbf{r},t)$ is the potential, $\Psi$ is the probability amplitude and the constants take on their usual meaning \cite{griffiths1995introduction}. For a free particle, $V(\mathbf{r},t)=0$, so we receive
\begin{equation}
-\frac{\hbar^2}{2m}\grad^2\Psi(\mathbf{r})=i\hbar \frac{\d}{\d t}\Psi(\mathbf{r}).
\label{eq:Scheq}
\end{equation}
We can make the ansatz
\begin{equation}
\Psi(\mathbf{r},t)=\psi(\mathbf{r})\me^{-\ii E t/\hbar}
\end{equation} 
where $E$ is the energy of the particle. Here, $E/\hbar$ plays the role of the angular frequency $\omega$. Using this in \eqref{eq:Scheq}, we get
\begin{equation}
\grad^2\psi(\mathbf{r})+\frac{2m E}{\hbar^2}\psi(\mathbf{r})=0
\end{equation}
If we make
\begin{equation}
k=\frac{\sqrt{2m E}}{\hbar},
\label{eq:qmk}
\end{equation}
we recover the Helmholtz equation,
\begin{equation}
\grad^2\psi(\mathbf{r})+k^2\psi(\mathbf{r})=0.
\label{eq:qmhelm}
\end{equation}
So the solutions to Helmholtz equation found in the optics (such as the Bessel or LG beam) can also apply to matter waves, where $k$ is the wavevector which now has a relation to mass and energy given by \eqref{eq:qmk}. Thus, we should theoretically be able to create a Bessel beam from matter waves.

\subsubsection{The Bessel beam}
Solving the Helmholtz equation \eqref{eq:qmhelm} in cylindrical coordinates produces the Bessel beam \cite{durnin1987diffraction}, given by
\begin{equation}
\psi(r,\phi,z)=J_l(k_r r) \me^{\ii k_z z}\me^{\ii l \phi},
\label{eq:bessel}
\end{equation}
where $J_l(k_r r)$ is the Bessel function of the first kind, $l$ is orbital angular momentum (OAM) number of the beam leading to an OAM of $l\hbar$ per photon \cite{allen2003optical}, and the wave vector $k$ is given by
\begin{equation}
k^2=k_r^2+k_z^2.
\label{eq:Besk}
\end{equation}

The Bessel beam is a non-diffracting beam in that it retains its spatial structure through propagation. It is also capable of self-healing \cite{gori1987bessel} in that it reforms shortly after an obstruction. A Bessel beam is non-normalizable in the transverse plane, as $\int_0^\infty r |J_l(r)|^2 dr=\infty$, but close approximations can be found, making the Bessel beam more a mathematical tool than an experimentally realizable beam.

\subsubsection{The Laguerre-Gauss beam \label{sec:lag}}

The paraxial approximation can be applied to the Helmholtz equation in cylindrical coordinates to find the Laguerre-Gauss (LG) beam \cite{GaussianBeams}, given by
\begin{equation}
\psi_{l,p}(r,\phi,z)=\frac{C_{l,p}}{w(z)}\left(\frac{r\sqrt{2}}{w(z)}\right)^{|l|}\me^{-\frac{r^2}{w^2(z)}}L_p^{|l|}\left(\frac{2 r^2}{w^2(z)}\right)\me^{\ii k\frac{r^2}{2 R(z)}}\me^{\ii l \phi}\me^{\ii k z}\me^{-\ii \varphi(z)},
\label{eq:LG}
\end{equation}
where $C_{l,p}$ is a normalization constant, $L_p^{|l|}$ are the generalised Laguerre polynomials, and the spot size $w(z)$, radius of curvature $R(z)$ and Gouy phase $\varphi(z)$ are all given by
\begin{align}
w(z)&=w_0\sqrt{1+\left(\frac{z}{z_R}\right)^2}, \label{eq:spot} \\
R(z)&=z\left[1+\left(\frac{z_R}{z}\right)^2\right], \\
\varphi(z)&=(|l|+2p+1)\arctan\left(\frac{z}{z_R}\right), \\
z_R&=\frac{\pi w_0^2}{\lambda}=\frac{k w_0^2}{2},
\end{align}
We can use the orthogonality of the Laguerre polynomials \cite{WolframLG} to show that the normalization constant $C_{l,p}$ is given by
\begin{equation}
C_{l,p}=\sqrt{\frac{2 p!}{\pi(p+|l|)!}}.
\end{equation}

\subsection{Analytic solution of the electromagnetic field surrounding a non-accelerating isotropic Bessel beam}\label{subsec:eBesselEM}

We wish to substitute the charge density $\rho$ for the probability distribution and the probability current $\jj$ of an electron beam that solves the Schr\"odinger equation into Maxwell's equations in order to solve for the electromagnetic $\E$ and $\B$ fields. Solving the resulting equations is outlined below.

\subsubsection{The electric field}

For a Bessel beam, the probability distribution, which results from solving the Schr\"odinger equation in cylindrical coordinates, is given by
\begin{equation}
\rho(r) =|J_l(k_r r)|^2.
\label{eq:SI_RhoBessel}
\end{equation}
In order to use this within Maxwell's equations we need to take into consideration the units of this quantity. In order to remain consistent, we need to multiply the above by a constant whose are are Coulombs per unit length, denoted by $\lambda$. Here the length is along the direction of the beam's propagation. Thus we use
\begin{equation}
\rho(r)=\lambda |J_l(k_r r)|^2
\end{equation}
with Gauss's law, to solve
\begin{equation}
\iint \E \cdot d\S=\frac{1}{\varepsilon_0}\iiint\rho(r) dV.
\end{equation}
Since we have an infinitely long cylindrically symmetric beam, the only direction $\E$ can face is $\E=E_r\rhat$. Thus, we construct a surface $\S$ which is a cylinder of radius $R$ and height $h$ that wraps around the propagation direction of the beam with an outward facing normal $\rhat$, i.e. $d\S=Rdzd\phi\rhat$. Gauss's Law then becomes
\begin{equation}
E_r 2\pi R h=\frac{1}{\varepsilon_0}\int_0^h dz\int_0^{2\pi}d\phi\int_0^R rdr \lambda |J_l(k_r r)|^2.
\end{equation}
Performing this integral and renaming $R$ to $r$, we get
\begin{equation}
E_r(r)=\frac{\lambda}{2\varepsilon_0 k_r}\left(k_r r|J_l(k_r r)|^2-2 l J_l(k_r r)J_{l-1}(k_r r)+k_r r |J_{l-1}(k_r r)|^2\right). \label{eq:BesselEr}
\end{equation}

\subsubsection{The magnetic field}
The probability current density for a Bessel beam is simply
\begin{equation}
\jj=\frac{\hbar}{m} |J_l(k_r r)|^2\left(\frac{l}{r}\phat+k_z\zhat\right).
\end{equation}
We want to use this of the electrical current density $\jj$. Again, in order to rectifiy the units of this quantity to remain consistent with that in Maxwell's equations, we must multiply $\jj$ by the same $\lambda$ as we did in the section above. We thus redefine $\jj$ as
\begin{equation}
\jj=\frac{\lambda \hbar}{m} |J_l(k_r r)|^2\left(\frac{l}{r}\phat+k_z\zhat\right).
\label{eq:BesselCurrent}
\end{equation}
For an infinitely long beam, we can use Amperian loops to determine the magnetic field. Note, the $\phat$ component of $\jj$ above is due solely to the OAM number $l$. This is not surprising, a beam carrying OAM is said to be a twisted beam, and the current density reflects this fact.

Since $\jj$ is independent of $\phi$ and $z$, we must have that $\B=\B(r)$. This implies that we must have the $r$ component of the magnetic field $B_r=0$. We can also prove that $B_r=0$ more explicitly by using the Biot-Savart law. Thus, the only components of $\B$ are
\begin{equation}
\B(r)=B_\phi(r)\phat+B_z(r)\zhat.
\end{equation}
We can construct two Amperian loops, namely a circle of radius $R$ centred about the origin denoted by $\d S_1$, and a square of height $h$ and length $R$ whose innermost side lies at $r=0$ denoted by $\d S_2$. So, Ampere's law, which is the the integral of \eqref{eq:Max4}, gives
\begin{equation}
\oint\B\cdot d\mathbf{l}=\mu_0 \iint \jj\cdot d\S,
\end{equation}
where $d\S=rdrd\phi\zhat$ for $\d S_1$ and $d\S=drdz\phat$ for $\d S_2$. Thus using Ampere's law and renaming $R$ to $r$ in the final answer we have for $\d S_1$
\begin{equation}
B_\phi(r)=\mu_0\frac{\lambda \hbar k_z}{2m k_r}\left(k_r r|J_l(k_r r)|^2-2 l J_l(k_r r)J_{l-1}(k_r r)+k_r r |J_{l-1}(k_r r)|^2\right),
\label{eq:BphiBes}
\end{equation}
and for $\d S_2$, we have
\begin{align}
B_z(r)=&B_z(0)-\mu_0\frac{l\lambda \hbar}{m}  4^{-|l|}(k_r r)^{2|l|}\Gamma(2|l|) \nonumber \\
&\times {}_p\tilde{F}_q\left(\left\{|l|,|l|+\frac{1}{2}\right\},\left\{|l|+1,|l|+1,2|l|+1\right\},-k_r^2 r^2\right),
\label{eq:BzBes}
\end{align}
where we have denoted ${}_p\tilde{F}_q(a,b,z)$ as the regularized generalized hypergeometric PFQ function \cite{WolframPFQ}. We can make the notation simpler by denoting
\begin{align}
\zeta_l(r)&=l 4^{-|l|}(k_r r)^{2|l|}\Gamma(2|l|) \nonumber \\
&\times {}_p\tilde{F}_q\left(\left\{|l|,|l|+\frac{1}{2}\right\},\left\{|l|+1,|l|+1,2|l|+1\right\},-k_r^2 r^2\right),
\end{align}
such that
\begin{equation}
B_z(r)=B_z(0)-\mu_0\frac{\lambda \hbar }{m}\zeta_l(r)
\end{equation}
To determine $B_z(0)$, we make use of the fact that we want $B_z(r)\to 0$ as $r\to \infty$. The second term reduces to a simple expression in this limit, namely \[\zeta_l(r)\to \frac{l}{2|l|}=\frac{1}{2}\sgn(l) \quad \text{as }r\to\infty.\] Thus,
\begin{equation}
B_z(0)=\mu_0\frac{\lambda \hbar	}{2m}\sgn(l),
\end{equation}
where $\sgn(l)$ gives the sign of $l$. We thus have
\begin{equation}
B_z(r)=\mu_0\frac{\lambda \hbar}{2m}\left(\sgn(l)-2\zeta_l(r)\right).
\end{equation}
We can denote $\lambda$ as the number of electrons per unit length multiplied by the charge $e$. So make $\lambda=\eta e$ where $\eta$ is the number of electrons per unit length. We thus have
\begin{gather}
B_\phi(r)=\frac{\eta\mu_0\mu_B k_z}{k_r}\left(k_r r|J_l(k_r r)|^2-2 l J_l(k_r r)J_{l-1}(k_r r)+k_r r |J_{l-1}(k_r r)|^2\right), \label{eq:BesselBphi}\\
B_z(r)=\eta\mu_0\mu_B\left(\sgn(l)-2\zeta_l(r)\right). \label{eq:BesselBz}
\end{gather}
Where
\begin{equation}
\mu_B=\frac{e\hbar}{2 m}=9.274\times 10^{-24}\rm J.T^{-1},
\end{equation}
is the Bohr Magneton.

\subsubsection{The full electromagnetic field solution}

A well known principle in quantum electrodynamics is that the components of the electric and magnetic field do not always commute, arising from the fact that the $\E$ field is related to an electron's position and the $\B$ field is related to its momentum. We've chosen to ignore these quantized field effects in our calculation and instead assume that our fields would be the result of an ensemble average over many independent measurements of the $\E$ and $\B$ field components. Thus, we treat the matter wave field quantum mechanically while treating the $\E$ and $\B$ fields classically.

Combining \eqref{eq:BesselEr}, \eqref{eq:BesselBphi}, and \eqref{eq:BesselBz}, we find that the full analytic solution to the electromagnetic field of an isotropic non-accelerating electron Bessel beam is
\begin{widetext}
\begin{align}
\E &= \frac{\eta e r}{2\varepsilon_0}\left[\left(J_\ell(k_r r)+J_{\ell-1}(k_r r)\right)^2-2\left(1+\frac{\ell}{k_r r}\right)J_\ell(k_r r)J_{\ell-1}(k_r r)\right]\rhat, \\
\B &= \eta \mu_0\mu_B k_z r\left[\left(J_\ell(k_r r)+J_{\ell-1}(k_r r)\right)^2-2\left(1+\frac{\ell}{k_r r}\right)J_\ell(k_r r)J_{\ell-1}(k_r r)\right]\phat \nonumber \\
&+ \eta \mu_0\mu_B\left[\sgn(\ell)-2 \ell (k_r r/2)^{2|\ell|}(2|\ell|-1)! {}_p\tilde{F}_q\left(\left\{|\ell|,|\ell|+\frac{1}{2}\right\},\left\{|\ell|+1,|\ell|+1,2|\ell|+1\right\},-k_r^2 r^2\right)\right]\zhat,
\end{align}
\end{widetext}
where $\eta$ is the number of electrons per unit longitudinal length, $e$ is the charge of the electron, $\varepsilon_0/\mu_0$ is the electric/magnetic vacuum permeability, $\mu_B$ is the Bohr magneton, and ${}_p\tilde{F}_q$ is the regularized generalized hypergeometric PFQ function \cite{WolframPFQ}.

\subsection{Analytic solution of the electromagnetic field surrounding a non-accelerating isotropic Laguerre-Gauss beam}

\subsubsection{The electric field}
The probability distribution for an LG beam, after fixing the units to match with an electrical charge density as done previously, is given by
\begin{equation}
\rho(r,z)=|\psi_{l,p}|^2=\frac{2 p!}{\pi (p+|l|)!}\frac{\lambda}{w^2(z)}\left(\frac{2 r^2}{w^2(z)}\right)^{|l|}\me^{-\frac{2 r^2}{w^2(z)}}\left[L_p^{|l|}\left(\frac{2 r^2}{w^2(z)}\right)\right]^2.
\label{eq:rhoLac}
\end{equation}
Now, $\rho$ is cylindrically symmetric and time-independent, so we would expect $\E$ to obey these symmetries and we also expect $E_\phi=0$, so $\E=E_r\rhat+E_z\zhat$. To calculate the electric field, let's define a cylindrical surface wrapped around the propagation direction of the beam, with radius $R$ and height $\Delta Z$ where $\Delta Z$ is small. The bottom cap will also be positioned at some $Z$ value. The idea here will be to consider varying orders of $\Delta z$ and make approximations to the true $\E$ field, where higher orders of $\Delta z$ lead to better approximations to $\E$. We thus have $d\S=Rd\phi dz\rhat$ for the cylindrical surface and $d\S=\pm rdrd\phi\zhat$ at the end caps. So,
\begin{align}
\iint \E\cdot d\S&=\iint\limits_{\rm cylinder}\E\cdot d\S\vert_{r=R} \nonumber \\
&+\iint\limits_{\text{bottom cap}} \E\cdot d\S \vert_{z=Z} \nonumber \\
&+ \iint\limits_{\text{top cap}} \E \cdot d\S \vert_{z=Z+\Delta Z}.
\end{align}
So,
\begin{align}
\iint \E\cdot d\S&=\int_0^{2\pi}d\phi\int_Z^{Z+\Delta Z}R E_r(R,z) \nonumber \\
&-\int_0^{2\pi}d\phi\int_0^Rrdr E_z(r,Z) \nonumber \\
&+\int_0^{2\pi}d\phi\int_0^Rrdr E_z(r,Z+\Delta Z).
\label{eq:LagESurf}
\end{align}
If $\Delta Z$ is small enough, then $E_r(R,z)$ is roughly constant over the range $Z$ to $Z+\Delta Z$. So we can approximate $E_r(R,z)\approx E_r(R,Z)$ and take it out of the integral to get
\begin{equation}
\int_0^{2\pi}d\phi\int_Z^{Z+\Delta Z}R E_r(R,z)\approx 2\pi R E_r(R,Z)\Delta Z.
\end{equation}
This approximation becomes exact in the limit $\Delta Z\to 0$. 

Now, the last two terms in \eqref{eq:LagESurf} are the same integral but over the difference $E_z(r,Z+\Delta Z)-E_z(r,Z)$. Using a Taylor series expansion on this term, we find
\begin{equation}
E(r,Z+\Delta Z)-E(r,Z)=\frac{\d}{\d z}E_z(r,Z)\Delta Z+\frac{1}{2}\frac{\d^2}{\d z^2}E_z(r,Z)\Delta Z^2+...
\end{equation}
Thus, \eqref{eq:LagESurf} becomes
\begin{align}
\iint \E\cdot d\S&\approx 2\pi R E_r(R,Z)\Delta Z \nonumber \\
&+2\pi \int_0^R r dr \left[\frac{\d}{\d z}E_z(r,Z)\Delta Z+\frac{1}{2}\frac{\d^2}{\d z^2}E_z(r,Z)\Delta Z^2+...\right].
\end{align}
This is the left hand side of Gauss's law. The right hand side has
\begin{equation}
\iint \E\cdot d\S=\frac{1}{\varepsilon_0}\int_0^R r dr\int_0^{2\pi} d\phi \int_Z^{Z+\Delta Z} dz \rho(r,z).
\end{equation}
Again, if $\Delta Z$ is small enough, then $\rho(r,z)$ can be treated as a constant within the integral, with $ \rho(r,z)\approx \rho(r,Z)$. Thus,
\begin{equation}
\frac{1}{\varepsilon_0}\int_0^R r dr\int_0^{2\pi} d\phi \int_Z^{Z+\Delta Z} dz \rho(r,z)\approx \frac{1}{\varepsilon_0} 2\pi \Delta Z\int_0^R r dr \rho(r,Z).
\end{equation}
Thus,
\begin{align}
\iint \E\cdot d\S&=2\pi R E_r(R,Z)\Delta Z \nonumber \\
&+2\pi\int_0^R r dr \left[\frac{\d}{\d z}E_z(r,Z)\Delta Z+\frac{1}{2}\frac{\d^2}{\d z^2}E_z(r,Z)\Delta Z^2+...\right] \nonumber \\
&=\frac{1}{\varepsilon_0}2\pi \Delta Z\int_0^R r dr \rho(r,Z)
\label{eq:LagEInts}
\end{align}
Taking the relevant orders of $\Delta Z$, we get
\begin{align}
\mathcal{O}\left(\Delta Z^1\right)&: R E_r(R,Z)+\int_0^R r dr \frac{\d}{\d z}E_z(r,Z) =\frac{1}{\varepsilon_0} \int_0^R r dr \rho(r,Z),  \label{eq:LagEDelZ1}\\
\mathcal{O}\left(\Delta Z^2\right)&: \int_0^R r dr \frac{\d^2}{\d z^2}E_z(r,Z)=0, \\
\mathcal{O}\left(\Delta Z^3\right)&: \int_0^R r dr \frac{\d^3}{\d z^3}E_z(r,Z)=0, \\
&\quad\quad\quad\quad\quad\quad \vdots \nonumber \\
\mathcal{O}\left(\Delta Z^n\right)&: \int_0^R r dr \frac{\d^n}{\d z^n}E_z(r,Z)=0. \quad \text{for }n\geq 2
\end{align}
We can solve all of these equations by setting
\begin{equation}
\frac{\d}{\d z}E_z(r,Z)=A+f(r).
\end{equation}
This implies
\begin{equation}
E_z(r,z)=Az+f(r)z+B,
\end{equation}
and if we want $E_z(r,z)\to 0$ as $z\to \pm \infty$ then we must have $A=B=0$ and $f(r)=0$. Using this, only \eqref{eq:LagEDelZ1} becomes non-trivial and we get
\begin{equation}
E_r(R,Z)=\frac{1}{\varepsilon_0 R}\int_0^R r dr \rho(r,Z).
\end{equation}
From here, we need to insert a value for the radial index $p$ into $\rho$ from \eqref{eq:rhoLac} in order to do the integral above. One of the more commonly used values for $p$ is $p=0$. This yields
\begin{equation}
E_r(r,z)=\frac{\lambda}{\varepsilon_0}\frac{|l|!-\Gamma\left(|l|+1,\frac{2 r^2}{w^2(z)}\right)}{2\pi r|l|!} \quad\quad (\text{for }p=0),
\label{eq:LagEp0}
\end{equation}
where $\Gamma(x,z)$ is the incomplete gamma function defined by
\begin{equation}
\Gamma(x,z)=\int_z^\infty t^{x-1}\me^{-t}dt.
\end{equation}
This actually satisfies $E_r(r,z)\to 0$ as $r\to \infty$, which tells us we are on the right track. In fact, $E_r(r,z)\to 0$ as $r\to \infty$ is satisfied for any radial index $p$. Thus all in all we have
\begin{equation}
\E=E_r(r,z)\rhat=\frac{\rhat}{\varepsilon_0 r}\int_0^r r' \rho(r',z)dr', \\
\label{eq:LagEint}
\end{equation}
where $\rho(r,z)$ is defined in \eqref{eq:rhoLac}.

\subsubsection{The magnetic field}
The current density for an LG beam is given by
\begin{equation}
\jj=\frac{\lambda \hbar}{m}\rho(r,z)\left(\frac{k r}{R(z)}\rhat+\frac{l}{r}\phat+\tilde{K}(r,z)\zhat\right)
\end{equation}
where $\rho$ is given in \eqref{eq:rhoLac} divided by $\lambda$, and
\begin{align}
\tilde{K}(r,z)= k-k\left(1-\left(\frac{z_R}{z}\right)^2\right)\frac{r^2}{2 R^2(z)}-(|l|+2p+1)\frac{z_R}{z^2+z_R^2},
\end{align}
which is the effective wavenumber.

To determine which components of $\B$ survive, we need to think carefully. Consider each component of the current density, and consider appropriate Amperian loops containing these components. The $\zhat$ components in $\jj$ can only give rise to $\rhat$ and $\phat$ components in $\B$ and if $j_z$ has azimuthal symmetry, then the $\rhat$ components in the magnetic field cancel. The $\phat$ components in $\jj$ can only produce $\rhat$ and $\zhat$ components in $\B$, and if $j_\phi$ does not depend on $z$ then the $\rhat$ components in $\B$ cancel once again (think of the solenoid). Finally, the $\rhat$ components in $\jj$ can only produce $\phat$ and $\zhat$ components in $\B$. 
\\ \\
Now, $j_z$ does not depend on $\phi$ (it has azimuthal symmetry) and $j_\phi$ does not depend on $z$, so the $\rhat$ terms in $\B$ cancel and we get
\begin{equation}
\B(r,z)=B_\phi(r,z)\phat+B_z(r,z)\zhat.
\end{equation}
So let's use Amperian loops as we did before to calculate the components of $\B$. The first loop, denoted $\d S_1$, is the edge of a disk at height $z$ of radius $r$. So $d\S=r'dr'd\phi\zhat$ for $\d S_1$. Ampere's law then yields,
\begin{align}
B_\phi(r,z)=&-\frac{\mu_0}{r}\frac{\lambda \hbar}{m}k\left(1-\left(\frac{z_R}{z}\right)^2\right)\frac{1}{2 R^2(z)}\int_0^r r'^3\rho(r',z)dr' \nonumber \\
&+\frac{\mu_0}{r}\frac{\lambda \hbar}{m}\left[k-(|l|+2p+1)\frac{z_R}{z^2+z_R^2}\right]\int_0^r r' \rho(r',z)dr'.
\label{eq:Lagbphiint}
\end{align}
The integrals look similar to those in \eqref{eq:LagEint}. 

We can solve for the other $\B$ components in a manner similar to that taken towards solving the electric field $\E$. Let's define a surface $\d S_2$ such that it starts at a height $Z$, it's inner edge lies at $r=0$ and it's outer edge lies at $r=R$. The surface's height is $\Delta Z$, where $\Delta Z$ is small. We also want it's unit normal to point in the $\phat$ direction, so $d\S=dr dz \phat$. If we perform this procedure and use the appropriate assumptions, we get 
\begin{equation}
B_z(r,z)=B_z(0,z)-\mu_0 l\frac{\lambda \hbar}{m}\int_0^r \frac{1}{r'} \rho(r',z)dr'.
\end{equation}
We find $B_z(0,z)$ by requiring $B_z(r,z)\to 0$ as $r\to \infty$. This yields
\begin{equation}
B_z(0,z)=\mu_0 l\frac{\lambda \hbar}{m}\int_0^\infty \frac{1}{r'} \rho(r',z)dr'.
\label{eq:LagBzint}
\end{equation}
Again, we must substitute a desired value for $p$ into the expression for $\rho$ before we can evaluate these integrals.

\end{document}